\documentclass[12pt,english,aps,nofootinbib,prb,preprint,superscriptaddress,showpacs,showkeys,12pt,sort&compress]{revtex4}
\usepackage{ae,aecompl}
\usepackage[T1]{fontenc}
\usepackage[latin9]{inputenc}
\usepackage{amsmath}
\usepackage{graphicx}
\PassOptionsToPackage{normalem}{ulem}
\usepackage{ulem}

\makeatletter

\providecommand{\tabularnewline}{\\}

\@ifundefined{textcolor}{}
{%
 \definecolor{BLACK}{gray}{0}
 \definecolor{WHITE}{gray}{1}
 \definecolor{RED}{rgb}{1,0,0}
 \definecolor{GREEN}{rgb}{0,1,0}
 \definecolor{BLUE}{rgb}{0,0,1}
 \definecolor{CYAN}{cmyk}{1,0,0,0}
 \definecolor{MAGENTA}{cmyk}{0,1,0,0}
 \definecolor{YELLOW}{cmyk}{0,0,1,0}
}

\usepackage{amsfonts}
\usepackage{aecompl}\usepackage{epstopdf}

\usepackage{babel}
\date{\today}

\usepackage{babel}

\makeatother

\usepackage{babel}
\begin{document}

\title{Phases, phase equilibria and phase rules in low-dimensional systems}

\author{\noindent T. Frolov}

\email{timfrol@berkeley.edu}

\address{Department of Materials Science and Engineering, University of California,
Berkeley, California 94720, USA}

\author{\noindent Y. Mishin }

\email{ymishin@gmu.edu}

\address{Department of Physics and Astronomy, MSN 3F3, George Mason University,
Fairfax, Virginia 22030, USA}
\begin{abstract}
We present a unified approach to thermodynamic description of one,
two and three dimensional phases and phase transformations among them.
The approach is based on a rigorous definition of a phase applicable
to thermodynamic systems of any dimensionality. Within this approach,
the same thermodynamic formalism can be applied for the description
of phase transformations in bulk systems, interfaces, and line defects
separating interface phases. For both lines and interfaces, we rigorously
derive an adsorption equation, the phase coexistence equations, and
other thermodynamic relations expressed in terms of generalized line
and interface excess quantities. As a generalization of the Gibbs
phase rule for bulk phases, we derive phase rules for lines and interfaces
and predict the maximum number of phases than may coexist in systems
of the respective dimensionality.
\end{abstract}


\maketitle

\section{Introduction\label{sec:Introduction}}

Surfaces and interfaces can affect many properties of materials ranging
from chemical reactivity to wetting, mechanical behavior and thermal
and electric resistance.\cite{Balluffi95} It has long been known
that interface properties can suddenly change due to an abrupt change
in their atomic structure and/or local chemical composition. Such
changes are often interpreted as transformations between different
interface phases and are usually described by well-established thermodynamic
theories of phase transformations. Despite many years of experimental
and theoretical studies, certain thermodynamic aspects of interface
phases and interface phase transformations remain unclear. In fact,
there is even a controversy about the thermodynamic nature of interface
phases, namely, whether they should be treated as a particular case
of the general concept of a phase in thermodynamics, or as something
fundamentally different from bulk phases. As a consequence, terminological
disagreements arose in the materials community, with some authors
referring to interface phases as ``phases''\cite{Hart:1972aa,Cahn82a,Pandit1982,Rottman1988a,Forgacs:1991aa,Straumal04,Mishin09c,Mishin2010a,Frolov2013,Frolov2013a,Frolov:2014aa}
while others prefer the new term ``complexion''\cite{Tang06,Tang06b,Dillon2007,Kaplan2013,Cantwell-2013,Zhou:2015aa}
introduced to avoid the association with bulk phases. 

The goal of this paper is to examine the parallel between phase transformations
in bulk and low-dimensional systems from the standpoint of classical
thermodynamics. Some of the questions that we seek to answer include:
To what extent can one apply the formalism and terminology of classical
three-dimensional (3D) thermodynamics to describe two-dimensional
(2D) phases at interfaces and one-dimensional (1D) phases within line
defects? Is it justifiable to treat transformations between 2D and
1D phases the same way as we treat transformations between bulk (3D)
phases? 

To answer these questions, we start by reviewing the original definition
of a phase,\cite{Willard_Gibbs} the modern definitions,\cite{Callen_book_1985,Tisza:1961aa}
and their applicability low-dimensional thermodynamic systems. We
then formulate a unified definition of a phase that spans all dimensionalities.
This definition identifies the concept of a phase with a fundamental
equation of state possessing a particular set of mathematical properties.
This unified definition leads to a unified treatment of phase equilibria
among phases of different dimensionality. Applying this unified thermodynamic
formalism, we rigorously derive 2D and 1D versions of the adsorption
equation, both expressed in terms of generalized interface (respectively,
line) excesses. We also derive generalized Clapeyron-Clausius type
equations describing the hypersurfaces of thermodynamic equilibrium
between bulk, interface or line phases, as well as generalized phase
rules for interfaces and line defects.

The approach applied in this work is largely influenced by the attempts
of many authors to formulate the logical structure of thermodynamics
in the form of axioms and postulates in a manner similar to mathematical
theories.\cite{Caratheodory_1909,Ehrenfest:1959aa,Tisza:1961aa,Landsberg:1970,Kestin:1970,Gurney:1970aa,Masavetas:1988aa,Jongschaap:2001aa,Callen_book_1985,Weihold_book}
It is not our goal to pursue a complete axiomatic structure of thermodynamics
in the present paper. However, the formulation of certain key points
in the form of definitions and postulates helps us emphasize important
concepts, logical connections and assumptions that are often implied
but not stated explicitly, or simply overlooked. The fundamental equations
of state are formulated for simple isotropic systems such as multi-component
fluids. While extensions to other systems are possible in the future,
in this paper we wish to focus the attention on the most basic concepts
and not be distracted by technical difficulties that arise in addressing
more complex systems.

As already mentioned, the approach proposed here is restricted to
classical thermodynamics. Thus, statistical-mechanical treatments
of phases of any dimensionality are beyond the scope of this paper.
The interested reader is referred to special literature in the respective
fields, such as the Rowlinson and Widom book\cite{Rawlinson1984}
on molecular-level theories of interfaces. Likewise, we refrain from
making any model assumptions regarding the local behavior of properties
inside low-dimensional phases. In particular, despite the success
of interface theories based on gradient thermodynamics\cite{Van_der_Waals,Cahn58a},
they are not part of the present discussion. Instead, we adhere to
the purely thermodynamic approach in which low-dimensional phases
are treated in terms of interface or line excess quantities without
attempting to characterize the excess regions in a more detailed (but
necessarily approximate) manner.

\section{Phases in bulk thermodynamics}

\subsection{Definition of a phase\label{sub:bulk_phase}}

The term ``phase'' was introduced by Gibbs,\cite{Willard_Gibbs} who
defined phases rather vaguely as ``different homogeneous bodies''
that ``differ in composition or state'' (Ref.~\onlinecite{Willard_Gibbs},
p.96). Despite the vagueness of this verbal definition, the equations
written by Gibbs make it clear that by a phase Gibbs understood a
continuum of spatially homogeneous thermodynamic states that follow
a given fundamental equation. The concept of a fundamental equation
was introduced by Gibbs 10 pages earlier (Ref.~\onlinecite{Willard_Gibbs},
p.86) and was defined as ``a single equation from which all these
relations {[}thermodynamic properties{]} may be deduced''. In most
of his work,\cite{Willard_Gibbs} Gibbs used the fundamental equation
expressing the entropy $S$ as a function of energy $U$, volume $V$,
and the amounts of $k$ chemical components $N_{1},...,N_{k}$ present
in the system. Thus, in modern notations, Gibbs' fundamental equation
has the form
\begin{equation}
S=S(U,V,N_{1},...,N_{k}).\label{eq:1}
\end{equation}
Occasionally, Gibbs used the function
\begin{equation}
U=U(S,V,N_{1},...,N_{k}),\label{eq:2}
\end{equation}
which is nowadays called the fundamental equation in the energy representation.\cite{Callen_book_1985}
The term ``fundamental'' expresses the important property of this
equation that, knowing it, all thermodynamic properties of the phase
can be derived by computing first and higher partial derivatives.

For solid phases, which were treated by Gibbs in a separate chapter,
the fundamental equations are more complex because the extensive properties
depend not only on the system volume but also on the elastic strain
relative to a chosen reference state. In addition, crystalline solids
are subject to a constraint on variations in the numbers of chemical
components imposed by the integrity of the crystalline lattice.\cite{Larche73,Larche_Cahn_78,Larche1985}
Here, we will limit the analysis to simple fundamental equations in
the form of Eqs.(\ref{eq:1}) and (\ref{eq:2}).

The association of phases with fundamental equations is also evident
from Gibbs' treatment of phase equilibria and phase transformations.\cite{Willard_Gibbs}
Fixing one of the extensive variables, say volume, the fundamental
equation can be rewritten in the density form, e.g.,
\begin{equation}
u=u(s,n_{1},...,n_{k}),\label{eq:2-1}
\end{equation}
where the small letters represent volume densities. The density form
reflects the fact that the identity of a phase does not depend on
its amount. Equation (\ref{eq:2-1}) can be represented by a hypersurface
in the space spanned by the density variables $(s,u,n_{1},...,n_{k})$,
sometimes referred to as the Gibbs space.\cite{Tisza:1961aa} Different
phases are represented by different hypersurfaces, which Gibbs called
the ``primitive surfaces''. Equilibria between different phases are
then described by imagining common tangent planes to the primitive
surfaces, the traces of their intersection with the primitive surfaces,
and other geometric constructions. Local curvature of the primitive
surface determines intrinsic thermodynamic stability of the respective
phase. Details of Gibbs' geometric thermodynamics\cite{Weihold_book}
will not be discussed here. The important point is that these constructions
imply an association of phases with different hypersurfaces and thus
different fundamental equations which define them.

After Gibbs, a number of authors re-examined the conceptual foundations
of the Gibbs' thermodynamics in pursuit of a more rigorous, axiomatic
structure of the discipline. The effort to create a formal structure
of thermodynamics started with the works of Carathéodory\cite{Caratheodory_1909}
and Ehrenfest\cite{Ehrenfest:1959aa} and continues to this day.\cite{Tisza:1961aa,Landsberg:1970,Kestin:1970,Gurney:1970aa,Masavetas:1988aa,Jongschaap:2001aa,Callen_book_1985,Weihold_book}
Probably, the most complete axiomatic formulation of thermodynamics
was developed by Tisza.\cite{Tisza:1961aa} We will take his approach
as a foundation for the present analysis. An important common feature
of all axiomatic structures of thermodynamics is the firm association
of the concept of phase with a fundamental equation. In essence, phases
are \emph{identified} with fundamental equations. Symbolically, 
\begin{equation}
\textnormal{Phase = Fundamental Equation.}\label{eq:0}
\end{equation}
We will, therefore, adopt the following definition of a bulk phase:

\textsl{\uline{Definition.}} Bulk phase is a set of spatially homogeneous
states of matter described by a given fundamental equation (\ref{eq:1})
with the following properties:
\begin{itemize}
\item $(U,V,N_{1},...,N_{k})$ are extensive (additive) parameters conserved
in any isolated system (``additive invariants'')\cite{Tisza:1961aa}
\item $S(U,V,N_{1},...,N_{k})$ is a homogeneous function of first degree
with respect to all arguments 
\item $S(U,V,N_{1},...,N_{k})$ is a smooth (infinitely differentiable)
function
\end{itemize}
The additivity of energy is only satisfied for short-range interatomic
forces. The requirement that $S(U,V,N_{1},...,N_{k})$ be a homogeneous
first degree function is critically important and implies scale invariance
of all thermodynamic properties of a phase. It is this property that
allows us to reformulate the fundamental equation in the density form.
The scale invariance breaks down near critical points.

According to the above definition, if two states of a system satisfy
the same fundamental equation, then thermodynamically, they represent
the same phase.

\subsection{Properties of a single phase}

All equilibrium thermodynamic properties of a single phase can be
derived from its fundamental equation by straightforward application
of calculus without any additional assumptions or approximations.
The calculations are simplified by using the following properties
of homogeneous functions.\cite{Courant:1989aa}

A function $f(x_{1},...,x_{n},y_{1},...,y_{m})$ is homogeneous of
degree one with respect to the variable set $(x_{1},...,x_{n})$ if
for any $\lambda>0$
\begin{equation}
f(\lambda x_{1},...,\lambda x_{n},y_{1},...,y_{m})=\lambda f(x_{1},...,x_{n},y_{1},...,y_{m}).\label{eq:3}
\end{equation}
For such functions, the Euler theorem states that
\begin{equation}
f(x_{1},...,x_{n},y_{1},...,y_{m})=\sum_{i=1}^{n}\dfrac{\partial f}{\partial x_{i}}x_{i}.\label{eq:4}
\end{equation}
Taking the full differential of this equation, we obtain
\begin{equation}
\sum_{j=1}^{m}\dfrac{\partial f}{\partial y_{j}}dy_{j}-\sum_{i=1}^{n}x_{i}d\left(\dfrac{\partial f}{\partial x_{i}}\right)=0.\label{eq:5}
\end{equation}
The presence of the non-scalable variables $y_{j}$ is optional. They
are needed in some applications.

We now apply Euler's theorem to the fundamental equation (\ref{eq:2})
with $x_{i}$ identified with $S$, $V$ and $N_{i}$ and without
the $y_{j}$-variables. We have
\begin{equation}
U=TS-pV+\sum_{i=1}^{k}\mu_{i}N_{i},\label{eq:6}
\end{equation}
where the temperature $T$, pressure $p$ and chemical potentials
$\mu_{i}$ are \emph{defined} by $T\equiv\partial U/\partial S$,
$p\equiv-\partial U/\partial V$ and $\mu_{i}\equiv\partial U/\partial N_{i}$,
respectively. Next, we apply Eq.(\ref{eq:5}) to obtain 
\begin{equation}
-SdT+Vdp-\sum_{i=1}^{k}N_{i}d\mu_{i}=0,\label{eq:7}
\end{equation}
which is the well-known the Gibbs-Duhem equation. The calculations
can be continued by computing higher derivatives of $U(S,V,N_{1},...,N_{k})$
to obtain the heat capacity, compressibility, thermal expansion factor
and all other commonly used thermodynamic properties. 

Equation (\ref{eq:7}) imposes a constraint on possible variations
of the $(k+2)$ intensive variables $(T,p,\mu_{1},...,\mu_{k})$ characterizing
the phase. Due to this constraint, any fixed amount of a single phase
is capable of
\begin{equation}
f=k+1\label{eq:8}
\end{equation}
independent variations called the thermodynamic degrees of freedom.

\subsection{Equilibrium in heterogeneous systems. The phase rule\label{sub:bul_phase_rule}}

To address heterogeneous systems, i.e., systems composed of multiple
phases, we introduce two postulates:

\textsl{\uline{Postulate 1.}} Any homogeneous substance can potentially
exist in multiple phases, each with its own fundamental equation. 

\textsl{\uline{Postulate 2.}} Any inhomogeneous substance is composed
of homogeneous subsystems representing phases.

Note that in the second Postulate, the subsystems can be either different
phases or different states of the same phase.\footnote{A homogeneous state is defined by a set of variables $(U,V,N_{1},...,N_{k})$.
Two homogeneous states represent the same phase if both satisfy the
same fundamental equation. }

Consider an isolated heterogeneous systems composed of $\varphi$
phases described by the fundamental equations
\begin{equation}
S_{n}(U_{n}V_{n},N_{n1},...,N_{nk}),\enskip n=1,...,\varphi.\label{eq:9}
\end{equation}
Neglecting, for right now, the role of interfaces between the phases,
the additivity of entropy dictates that the total entropy of the system
is
\begin{equation}
S_{tot}=\sum_{n}S_{n}(U_{n}V_{n},N_{n1},...,N_{nk}).\label{eq:10}
\end{equation}
According to the entropy maximum principle,\cite{Willard_Gibbs,Tisza:1961aa,Callen_book_1985}
the necessary and sufficient condition of equilibrium of the system
is the maximum of $S_{tot}$ under the isolation constraints\cite{Willard_Gibbs}
\begin{equation}
\sum_{n}U_{n}=\textrm{const},\label{eq:11}
\end{equation}
\begin{equation}
\sum_{n}V_{n}=\textrm{const},\label{eq:12}
\end{equation}
\begin{equation}
\sum_{n}N_{ni}=\textrm{const},\enskip i=1,...,k,\label{eq:13}
\end{equation}
expressing the conservation of energy, volume and the amount of each
chemical component, respectively. This variational problem is solved
by using a set of Lagrange multipliers $\lambda_{u}$,$\lambda_{v}$
and $\lambda_{i}$,
\begin{equation}
\sum_{n}S_{n}-\lambda_{u}\left(\sum_{n}U_{n}\right)-\lambda_{v}\left(\sum_{n}V_{n}\right)-\sum_{i}\lambda_{i}\left(\sum_{n}N_{ni}\right)\rightarrow\textrm{max}.\label{eq:14}
\end{equation}
The well-known solution is the equality of temperatures, pressures
and chemical potentials in all phases:
\begin{equation}
T_{1}=...=T_{\varphi}\equiv T,\label{eq:15}
\end{equation}
\begin{equation}
p_{1}=...=p_{\varphi}\equiv p,\label{eq:16}
\end{equation}
\begin{equation}
\mu_{1i}=...=\mu_{\varphi i}\equiv\mu_{i},\enskip i=1,...,k.\label{eq:17}
\end{equation}

Thus the entire heterogeneous system is described by $(k+2)$ intensive
variables $(T,p,\mu_{1},...,\mu_{k})$. However, these variables are
not independent. Each phase must satisfy its own Gibbs-Duhem equation,
which imposed $\varphi$ constraints
\begin{equation}
-S_{n}dT+V_{n}dp-\sum_{i=1}^{k}N_{ni}d\mu_{i}=0,\enskip n=1,...,\varphi.\label{eq:18}
\end{equation}
As a result, the number of independent variations of the heterogeneous
system becomes
\begin{equation}
f=k+2-\varphi,\label{eq:19}
\end{equation}
a relation known as the Gibbs phase rule.\cite{Willard_Gibbs}

The global properties $(T,p,\mu_{1},...,\mu_{k})$ are referred to
as intensities or fields.\cite{Griffiths-Wheeler-1970} They are distinguished\cite{Griffiths-Wheeler-1970}
from densities (ratios of extensive properties), such as energy, entropy,
or the amounts of chemical components per unit volume or per particle.
Both intensities and densities are local variables that characterize
physical points. However, while the intensities are uniform across
the heterogeneous system, the densities are generally different in
different phases and experience discontinuities across phase boundaries.
The geometric common-tangent constructions describing phase equilibria\cite{Willard_Gibbs,Weihold_book}
apply to densities but not intensities.

\section{Interface phases}

\subsection{Definition of interface phase}

The concept of an interface phase can be traced back to Gibbs.\cite{Willard_Gibbs}
When analyzing the stability of interfaces with respect to changes
in state (Ref.~\onlinecite{Willard_Gibbs}, p.~237-240), Gibbs recognized
the possibility of different interface states that can reach equilibrium
with the same bulk phases. Gibbs did not call these equilibrium states
phases, but he treated them exactly the same way as he treated bulk
phases in other parts of his work.\cite{Willard_Gibbs} He showed
that if the interface ``states'' coexist in equilibrium with each
other, they must have the same surface tension $\gamma$. He also
showed that interface states with lower $\gamma$ are more stable
than interface states with larger $\gamma$. In other words, the most
stable state of the interface is that which minimizes the interface
tension. Gibbs even discussed metastable states of interfaces and
pointed out that they can transform to more stable states by a nucleation
and growth mechanism.

Later on, the interface ``states'' discussed by Gibbs came to be called
surface or interface phases,\cite{Cahn82a,Pandit1982,Rottman1988a,Straumal04}
especially in the surface physics and chemistry communities where
a large variety of surface phases were found in adsorbed surface layers
and represented as surface phase diagrams.\cite{King:1994aa} In the
1970s, Hart published influential papers analyzing structural transformations
in grain boundaries.\cite{Hart:1972aa} Hart applied the thermodynamic
formalism developed by Gibbs,\cite{Willard_Gibbs} except that he
referred to Gibbs' interface ``states'' as 2D grain boundary phases.
Furthermore, Hart explicitly associated the grain boundary phases
with different fundamental equations. Using the analogy with Gibbs'
bulk thermodynamics, he derived a generalized version of the Clapeyron-Clausius
equation that contained jumps of interface excess properties between
the grain boundary phases, including a jump of the excess volume.
Cahn\cite{Cahn82a} published a thorough thermodynamic analysis and
the most complete classification of interface phase transformations.
Over the recent years, experiments have revealed a number of phases
and phase transformations in both metallic and ceramic grain boundaries,
see Refs.~\onlinecite{Cantwell-2013,Kaplan2013} for recent reviews.
Atomistic computer simulations have reached the stages where reversible
structural phase transformation can be identified and studied in both
low\cite{Olmsted2011} and high-angle\cite{Frolov2013,Frolov2013a}
grain boundaries.

\begin{figure}
\noindent \begin{centering}
\includegraphics[clip,width=0.75\textwidth]{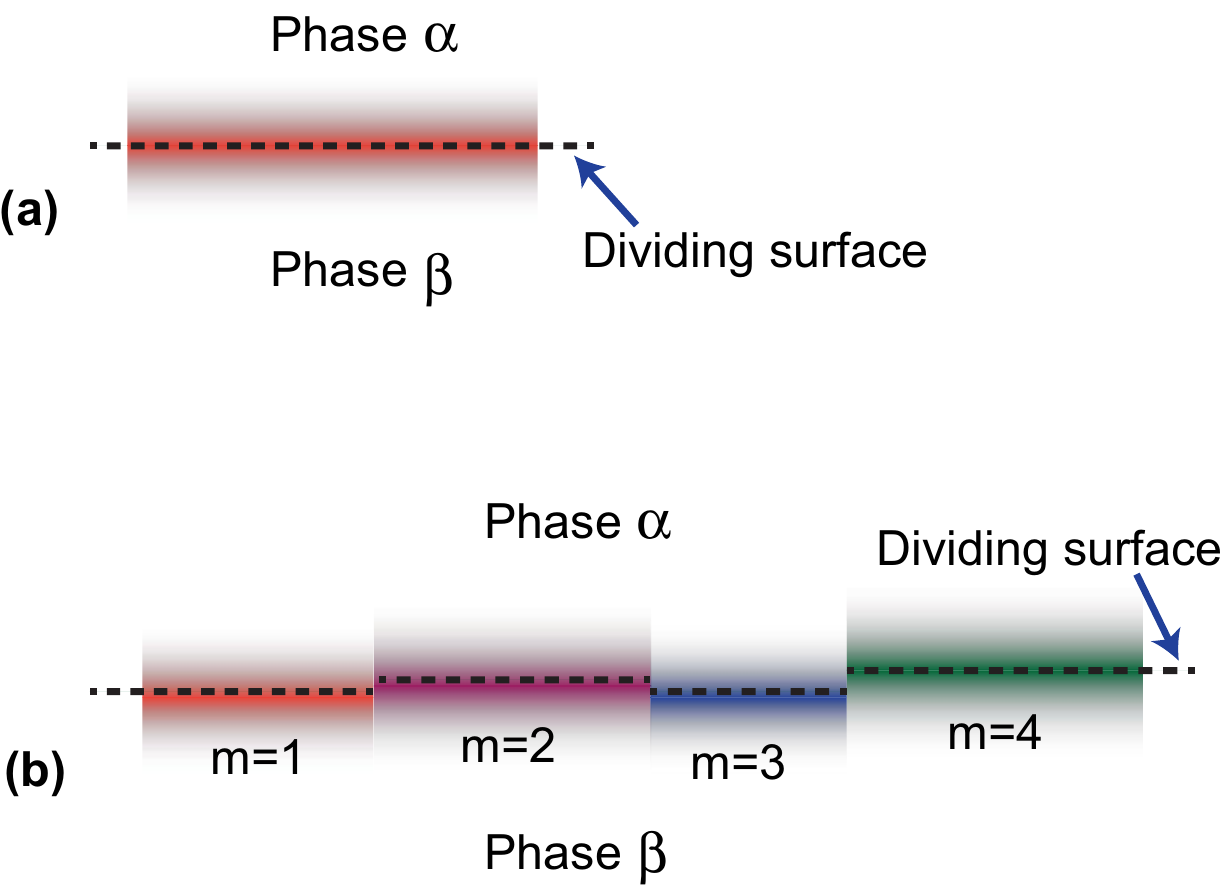}
\par\end{centering}

\protect\caption{(a) Single-phase interface between bulk phases $\alpha$ and $\beta$.
(b) Interface between the same bulk phases composed of $\nu=4$ interface
phases numbered by index $m$. The dividing surfaces are indicated
by dashed lines.\label{fig:interface_phases}}
\end{figure}

To formulate a rigorous definition of an interface phase, consider
two bulk phases, $\alpha$ and $\beta$, separated by a plane interface
(Fig.~\ref{fig:interface_phases}(a)). Following Gibbs,\cite{Willard_Gibbs}
we choose a geometric dividing surface by some rule. For example,
it can be the equimolar surface of component 1 (zero excess of this
component). This choice is unimportant and is only needed as a starting
point. We will soon replace it by a general formulation that does
not require a dividing surface. For any extensive property $X$, we
define its excess $\tilde{X}$ by
\begin{equation}
\tilde{X}=X-X_{\alpha}-X_{\beta}.\label{eq:20}
\end{equation}
Here $X$ is the value of the property for an imaginary box containing
the interface and $X_{\alpha}$ and $X_{\beta}$ are the values assigned
to the phases assuming that they remain homogeneous all the way to
the dividing surface.

\textsl{\uline{Definition.}} Interface phase is a set of spatially
homogeneous (over the dividing surface) states of the interface described
by a given fundamental equation 
\begin{equation}
\tilde{S}=\tilde{S}(\tilde{U},A,\tilde{N}_{2},...,\tilde{N}_{k})\label{eq:21}
\end{equation}
 with the following properties:
\begin{itemize}
\item $(\tilde{S},\tilde{U},A,\tilde{N}_{2},...,\tilde{N}_{k})$ are extensive
(additive) parameters on the dividing surface
\item $\tilde{S}(\tilde{U},A,\tilde{N}_{2},...,\tilde{N}_{k})$ is a homogeneous
function of first degree with respect to the variable set $(\tilde{U},A,\tilde{N}_{2},...,\tilde{N}_{k})$ 
\item $\tilde{S}(\tilde{U},A,\tilde{N}_{2},...,\tilde{N}_{k})$ is a smooth
(infinitely differentiable) function
\end{itemize}
Here, $A$ is the area of the dividing surface. The excesses of the
components, $\tilde{N}_{i}$, represent interface segregations or
depletions. Note that $\tilde{N}_{1}$ is not listed as a variable
due to our choice of the dividing surface ($\tilde{N}_{1}=0$). This
definition is similar to the bulk phase definition (Sec.~\ref{sub:bulk_phase}),
except that the volume is replaced by the area and the spatial homogeneity
is understood in the 2D sense (over the surface). While there can
be situations in which the excess properties display strong variations
along the interface, the present theory is limited to systems in which
gradients along the interface can be neglected. Small variations can
be easily handled by mentally partitioning the interface into regions
that can be treated as homogeneous with sufficient accuracy. Of course,
most properties exhibit extremely rapid spatial variations across
the interface. As already mentioned, the present theory does not set
the goal of describing such local variations but instead focuses on
the total, integrated amounts of the respective properties and their
excesses over bulk properties. 

Note that, similar to bulk phases, we postulate a one-to-one mapping
between interface phases and fundamental equations: different interface
phases - different fundamental equations.

It is important to note that the very existence of the interface fundamental
equation implies that interfaces can exist in states of internal equilibrium
that follow a fundamental equation without being in equilibrium with
the bulk phases or other interface phases. Recall that the same is
implied in bulk thermodynamics: bulk phases satisfy their fundamental
equations whether or not they are in equilibrium with each other.

\subsection{Equilibrium among interface phases.}

To describe multiple interface phases, we introduce the following
postulates:

\textsl{\uline{Postulate 1.}} Any homogeneous interface can potentially
exist in multiple interface phases, each with its own fundamental
equation. 

\textsl{\uline{Postulate 2.}} Any inhomogeneous interface is composed
of homogeneous regions representing interface phases.

As with bulk phases, the homogeneous regions mentioned in Postulate
2 can be either different interface phases or different states of
the same interface phase.

We next derive the conditions of thermodynamic equilibrium between
different interface phases and between the interface and the adjoining
bulk phases. Let the interface be composed of $\nu$ phases occupying
different areas $A_{m}$ (Fig.~\ref{fig:interface_phases}(b)). Their
fundamental equations are
\begin{equation}
\tilde{S}_{m}=\tilde{S}_{m}(\tilde{U}_{m},A_{m},\tilde{N}_{m2},...,\tilde{N}_{mk}),\enskip m=1,...,\nu.\label{eq:22}
\end{equation}
Note that the dividing surfaces of the interface phases are generally
different. To find the conditions of equilibrium, we apply the entropy
maximum principle\cite{Willard_Gibbs,Tisza:1961aa,Callen_book_1985}
with respect to the total entropy of the system considered to be isolated.
The calculation is similar to the bulk case (Sec.~\ref{sub:bul_phase_rule}),
except that we now include the contributions of interface phases that
were previously ignored. The total entropy combines the fundamental
equations of the bulk and interface phases: 
\begin{equation}
S_{tot}=S_{\alpha}(U_{\alpha}V_{\alpha},N_{\alpha1},...,N_{\alpha k})+S_{\beta}(U_{\beta}V_{\beta},N_{\beta1},...,N_{\beta k})+\sum_{m=1}^{\nu}\tilde{S}_{m}(\tilde{U}_{m},A_{m},\tilde{N}_{m2},...,\tilde{N}_{mk}).\label{eq:23}
\end{equation}
This entropy is maximized at fixed values of the total energy, volume,
interface area and the amounts of all chemical components:
\begin{equation}
U_{\alpha}+U_{\beta}+\sum_{m=1}^{\nu}\tilde{U}_{m}=\textrm{const},\label{eq:24}
\end{equation}
\begin{equation}
V_{\alpha}+V_{\beta}=\textrm{const},\label{eq:25}
\end{equation}
\begin{equation}
\sum_{m=1}^{\nu}A_{m}=\textrm{const},\label{eq:26}
\end{equation}
\begin{equation}
N_{\alpha i}+N_{\beta i}+\sum_{m=1}^{\nu}\tilde{N}_{mi}=\textrm{const},\enskip i=1,...,k.\label{eq:27}
\end{equation}
These constraints are imposed by Lagrange multipliers $\lambda_{u}$,
$\lambda_{v}$, $\lambda_{a}$ and $\lambda_{i}$, reducing the variational
problem to
\begin{eqnarray}
 &  & S_{\alpha}+S_{\beta}+\sum_{m=1}^{\nu}\tilde{S}_{m}-\lambda_{u}\left(U_{\alpha}+U_{\beta}+\sum_{m=1}^{\nu}\tilde{U}_{m}\right)-\lambda_{v}\left(V_{\alpha}+V_{\beta}\right)\nonumber \\
 & - & \lambda_{a}\left(\sum_{m=1}^{\nu}A_{m}\right)-\sum_{i}\lambda_{i}\left(N_{\alpha i}+N_{\beta i}+\sum_{m=1}^{\nu}\tilde{N}_{mi}\right)\rightarrow\textrm{max}.\label{eq:28}
\end{eqnarray}
The following equilibrium conditions are obtained:
\begin{itemize}
\item Thermal equilibrium
\begin{equation}
T_{\alpha}=T_{\beta}=\left(\dfrac{\partial\tilde{S}_{1}}{\partial\tilde{U}_{1}}\right)^{-1}=...=\left(\dfrac{\partial\tilde{S}_{\nu}}{\partial\tilde{U}_{\nu}}\right)^{-1}\equiv T.\label{eq:29}
\end{equation}

\item Mechanical equilibrium
\begin{equation}
p_{\alpha}=p_{\beta}.\label{eq:30}
\end{equation}

\item Chemical equilibrium
\begin{equation}
\mu_{\alpha i}=\mu_{\beta i}=-T\left(\dfrac{\partial\tilde{S}_{1}}{\partial\tilde{N}_{1i}}\right)=...=-T\left(\dfrac{\partial\tilde{S}_{\nu}}{\partial\tilde{N}_{\nu i}}\right)\equiv\mu_{i}\enskip i=2,...,k,\label{eq:31}
\end{equation}
\begin{equation}
\mu_{\alpha1}=\mu_{\beta1}.\label{eq:31a}
\end{equation}

\item Interface tension equilibrium
\begin{equation}
\dfrac{\partial\tilde{S}_{1}}{\partial A_{1}}=...=\dfrac{\partial\tilde{S}_{\nu}}{\partial A_{\nu}}.\label{eq:32}
\end{equation}

\end{itemize}
The derivatives $(\partial\tilde{S}_{m}/\partial\tilde{U}_{m})^{-1}$
can be called the ``temperatures'' of the interface phases, in which
case Eq.(\ref{eq:29}) expresses the uniformity of temperature across
the entire equilibrium system, including the interface and bulk phases.
Likewise, $-T(\partial\tilde{S}_{m}/\partial\tilde{N}_{mi})$ can
be called the ``chemical potential'' of component $i$ in the interface
phase $m$.\footnote{The interface chemical potential of component 1 is not defined by
Eqs.(\ref{eq:31}) and (\ref{eq:31a}) since its interface excess
is zero. However, we can repeat the calculation by choosing the dividing
surface as equimolar with respect to another component. This will
lead to the expected result that $\mu_{1}=\mu_{\alpha i}=\mu_{\beta i}$.} The chemical equilibrium can be then formulated as equality of chemical
potentials $\mu_{i}$ in all interface and bulk phases of the system.
Finally, 
\begin{equation}
\gamma_{m}\equiv-T\dfrac{\partial\tilde{S}_{m}}{\partial A_{m}}\label{eq:33}
\end{equation}
is defined as the interface free energy or interface tension of phase
$m$. Since we have postulated that the excess entropy is a homogeneous
function of first degree with respect to the area, it follows that
$\gamma_{m}$ is a local property that does not depend on the area.
For small interface regions bounded by other defects this may not
be the case. Such situations are not considered in the present theory. 

Thus, the interface tension equilibrium can be expressed as the equality
of tensions in all coexisting interface phases:
\begin{equation}
\gamma_{1}=...=\gamma_{\nu}.\label{eq:34}
\end{equation}
This equation recovers the interface equilibrium condition derived
by Gibbs.\cite{Willard_Gibbs} Just like the interface area is a 2D
analog of volume, the interface tension $\gamma$ is a 2D analog of
the bulk pressure $-p$. In this sense, Eq.(\ref{eq:32}) is a 2D
analog of the equality of pressures in coexisting bulk phases.

\subsection{Equilibrium properties of a single interface phase }

Similar to bulk phases, the fundamental equation (\ref{eq:21}) encapsulates
all properties of the interface phase. Consider reversible processes
in which a single-phase interface always remains in equilibrium with
the bulk phases. Since the fundamental equation is homogeneous first
degree, we apply Euler\textquoteright s theorem to obtain
\begin{eqnarray}
\tilde{S} & = & \dfrac{\partial\tilde{S}}{\partial\tilde{U}}\tilde{U}+\dfrac{\partial\tilde{S}}{\partial A}A+\sum_{i=2}^{k}\dfrac{\partial\tilde{S}}{\partial\tilde{N}_{i}}\tilde{N}_{i}\nonumber \\
 & = & \dfrac{1}{T}\tilde{U}-\dfrac{\gamma}{T}A-\dfrac{1}{T}\sum_{i=2}^{k}\mu_{i}\tilde{N}_{i},\label{eq:35}
\end{eqnarray}
which can be rewritten as
\begin{equation}
\gamma A=\tilde{U}-T\tilde{S}-\sum_{i=2}^{k}\mu_{i}\tilde{N}_{i}.\label{eq:36}
\end{equation}
This equation appears in Gibbs (Ref.~\onlinecite{Willard_Gibbs},
Eq.~(502)) and expresses $\gamma$ as the excess of the grand potential
$\tilde{U}-T\tilde{S}-\Sigma_{i}\mu_{i}\tilde{N}_{i}$ per unit interface
area. On the other hand, differentiation of the fundamental equation
(\ref{eq:21}) gives
\begin{equation}
d\tilde{S}=\dfrac{1}{T}d\tilde{U}-\dfrac{\gamma}{T}dA-\dfrac{1}{T}\sum_{i=2}^{k}\mu_{i}d\tilde{N}_{i},\label{eq:37}
\end{equation}
from which
\begin{equation}
d\tilde{U}=Td\tilde{S}+\sum_{i=2}^{k}\mu_{i}d\tilde{N}_{i}+\gamma dA.\label{eq:38}
\end{equation}
This well-known equation also appears in Gibbs (Ref.~\onlinecite{Willard_Gibbs},
Eq.~(501)) and shows that the interface adds an extra work term which
is a 2D analog of the mechanical work $-pdV$ in bulk systems. I fact,
this equation was the starting point of Gibbs' interface thermodynamics
from which all other equations were derived. Finally, by adding Eq.(\ref{eq:38})
to the differential of Eq.(\ref{eq:36}) we obtain the Gibbs adsorption
equation (Ref.~\onlinecite{Willard_Gibbs}, Eq.~(508))
\begin{equation}
Ad\gamma=-\tilde{S}dT-\sum_{i=2}^{k}\tilde{N}_{i}d\mu_{i}.\label{eq:39}
\end{equation}
Note the ease with which these equations have been derived starting
from the fundamental equation and using known properties of homogeneous
functions.

Equation (\ref{eq:39}) shows that the state of a single-phase interface
that maintains equilibrium with the bulk phases is defined by $k$
independent intensive variables. This is consistent with the Gibbs
phase rule (\ref{eq:19}) predicting $f=k$ degrees of freedom for
a two-phase ($\varphi=2$) system with $k$ chemical components.

\subsection{Reformulation in generalized interface excesses\label{sub:Reformulation-interfaces}}

Until this point, the interface excess quantities were defined relative
to a certain choice of the dividing surface. We can now remove this
restriction and reformulate all equations in terms of generalized
excess introduced by Cahn.\cite{Cahn79a} We start with the adsorption
equation (\ref{eq:39}) and replace all excesses appearing in this
equation by their definitions according to Eq.(\ref{eq:20}). As a
result, the adsorption equation takes the ``global'' form
\begin{equation}
Ad\gamma=-SdT+Vdp-\sum_{i=1}^{k}N_{i}d\mu_{i},\label{eq:40}
\end{equation}
where $S$, $V$ and $N_{i}$ refer to the entire system containing
two bulk phases and the interface. The terms with $X_{\alpha}$ and
$X_{\beta}$ canceled out due to the Gibbs-Duhem equations for the
bulk phases:
\begin{equation}
0=-S_{\alpha}dT+V_{\alpha}dp-\sum_{i=1}^{k}N_{\alpha i}d\mu_{i},\label{eq:41}
\end{equation}
\begin{equation}
0=-S_{\beta}dT+V_{\beta}dp-\sum_{i=1}^{k}N_{\beta i}d\mu_{i}.\label{eq:42}
\end{equation}
Since these equations remain valid after re-scaling by an arbitrary
factor, they can be thought of as representing arbitrarily chosen
homogeneous regions inside the bulk phases. Note that the global adsorption
equation (\ref{eq:40}) does not contain excess quantities, which
demonstrates that $\gamma$ is independent of definitions of excesses.

As a result of this ``unwrapping'' procedure, the adsorption equation
(\ref{eq:39}) has been recast in the global form (\ref{eq:40}) where
it does not depend on any definitions of interface excesses. However,
this equation must be considered simultaneously with the Gibbs-Duhem
equations (\ref{eq:43}) and (\ref{eq:42}). The advantage of this
global form is that we can now eliminate \emph{any} two differentials
from Eq.(\ref{eq:40}), not necessarily $dp$ and $d\mu_{1}$ as it
was done before by choosing the equimolar dividing surface of component
1. The elimination is accomplished most elegantly by using the Kramer
rule of linear algebra,\cite{Cahn79a} resulting in the generalized
adsorption equation:
\begin{equation}
Ad\gamma=-\left[S\right]_{XY}dT+\left[V\right]_{XY}dp-\sum_{i=1}^{k}\left[N_{i}\right]_{XY}d\mu_{i}.\label{eq:43}
\end{equation}
Here, $X$ and $Y$ are any two extensive variables from the list
$(S,V,N_{1},...,N_{k})$ and the square brackets denote rations of
two determinants. Namely, for any extensive property $Z$,
\begin{equation}
\left[Z\right]_{XY}\equiv\dfrac{\left\vert \begin{array}{ccc}
Z & X & Y\\
Z_{\alpha} & X_{\alpha} & Y_{\alpha}\\
Z_{\beta} & X_{\beta} & Y_{\beta}
\end{array}\right\vert }{\left\vert \begin{array}{cc}
X_{\alpha} & Y_{\alpha}\\
X_{\beta} & Y_{\beta}
\end{array}\right\vert },\label{eq:44}
\end{equation}
where $Z$, $X$ and $Y$ are computed for the two-phase system and
the remaining quantities represent arbitrarily chosen homogeneous
regions inside the bulk phases. The generalized excess $[Z]_{XY}$
has the meaning of the interface excess of the extensive property
$Z$ in a two-phase system that contains the same amounts of $X$
and $Y$ as the two single-phase regions combined. By properties of
determinants, $[X]_{XY}=[Y]_{XY}=0$. Thus, the excesses of $X$ and
$Y$ are zero and two terms in Eq.(\ref{eq:43}) automatically vanish.
By choosing the properties $X$ and $Y$ we can control which two
differentials in Eq.(\ref{eq:43}) disappear and which $k$ remain
as independent variables. 

If $X$ is volume, then $[Z]_{XY}$ defines an excess of $Z$ relative
to a geometric dividing surface as in Gibbs' interface thermodynamics.\cite{Willard_Gibbs}
The position of the dividing surface is then dictated by the choice
of the second variable $Y$. For example, by choosing $Y=N_{1}$ we
return to the previously defined excesses relative to the equimolar
surface of component 1. When neither $X$ nor $Y$ is volume, the
generalized interface excess is defined without using any dividing
surface. This generalization enables us to define the interface excess
volume $[V]_{XY}$. As all other excesses, the excess volume is not
unique. For example, if we choose $X=S$ and $Y=N_{1}$, the excess
volume becomes 
\begin{equation}
\left[V\right]_{SN_{1}}=\dfrac{\left\vert \begin{array}{ccc}
V & S & N_{1}\\
V_{\alpha} & S_{\alpha} & N_{\alpha1}\\
V_{\beta} & S_{\beta} & N_{\beta1}
\end{array}\right\vert }{\left\vert \begin{array}{cc}
S_{\alpha} & N_{\alpha1}\\
S_{\beta} & N_{\beta1}
\end{array}\right\vert }.\label{eq:45}
\end{equation}

Other interface properties can also be reformulated in terms for generalized
excesses. For example, the interface tension equation (\ref{eq:36})
can be generalized to\cite{Frolov09b} 
\begin{equation}
\gamma A=\left[U\right]_{XY}-T\left[S\right]_{XY}+p\left[V\right]_{XY}-\sum_{i=1}^{k}\mu_{i}\left[N_{i}\right]_{XY}.\label{eq:46}
\end{equation}
This leads to several equivalent excess formulations of $\gamma$,
such as 
\begin{equation}
\gamma A=\left[U+pV-\sum_{i=2}^{k}N_{i}\mu_{i}\right]_{SN_{1}}=\left[U-TS+pV-\sum_{i=3}^{k}N_{i}\mu_{i}\right]_{N_{1}N_{2}}.\label{eq:47}
\end{equation}

We can also derive an interface analog of the Clapeyron-Clausius equation
describing phase coexistence. Indeed, for two coexisting interface
phase, labeled by a single and double prime, Eq.(\ref{eq:43}) takes
the form

\begin{equation}
d\gamma^{\prime}=-\dfrac{\left[S^{\prime}\right]_{XY}}{A^{\prime}}dT+\dfrac{\left[V^{\prime}\right]_{XY}}{A^{\prime}}dp-\sum_{i=1}^{k}\dfrac{\left[N_{i}^{\prime}\right]_{XY}}{A^{\prime}}d\mu_{i},\label{eq:43-1}
\end{equation}

\begin{equation}
d\gamma^{\prime\prime}=-\dfrac{\left[S^{\prime\prime}\right]_{XY}}{A^{\prime\prime}}dT+\dfrac{\left[V^{\prime\prime}\right]_{XY}}{A^{\prime\prime}}dp-\sum_{i=1}^{k}\dfrac{\left[N_{i}^{\prime\prime}\right]_{XY}}{A^{\prime\prime}}d\mu_{i}.\label{eq:43-2}
\end{equation}
In reversible processes when the phases remain in equilibrium with
each other, their tensions must remain equal, $d\gamma^{\prime}=d\gamma^{\prime\prime}$.
This immediately gives
\begin{equation}
-\Delta\left(\dfrac{\left[S\right]_{XY}}{A}\right)dT+\Delta\left(\dfrac{\left[V\right]_{XY}}{A}\right)dp-\sum_{i=1}^{k}\Delta\left(\dfrac{\left[N_{i}\right]_{XY}}{A}\right)d\mu_{i}=0,\label{eq:43-3}
\end{equation}
where $\Delta$ denotes the difference between the two interface phases.
For example, by choosing $X=S$ and $Y=V$, we obtain the Clapeyron-Clausius
type equation
\begin{equation}
-\sum_{i=1}^{k}\Delta\left(\dfrac{\left[N_{i}\right]_{XY}}{A}\right)d\mu_{i}=0\label{eq:43-4}
\end{equation}
defining a $(k-1)$-dimensional hypersurface of interface phase coexistence
in the $k$-dimensional space of variables $(\mu_{1},...,\mu_{k})$.
The coefficients in Eq.(\ref{eq:43-4}) are the jumps of interface
segregations (per unit area) across this hypersurface.

\subsection{The interface phase rule }

We can now formulate the interface phase rule. Consider an equilibrium
heterogeneous system composed of $\varphi$ coexisting bulk phases.
We will focus the attention on one particular interface separating
two phases. Suppose this interface contains $\nu$ coexisting interface
phases. We wish to find the number of independent variables that can
be varied while keeping the same number of bulk and interface phases.
Let us call this number $f_{i}$ degrees of freedom.

The system is described by $(k+2)$ intensive parameters: $T$, $p$,
and $k$ chemical potentials $\mu_{i}$. But they are not independent.
First, we have $\varphi$ Gibbs-Duhem equations (\ref{eq:18}) for
the bulk phases that impose $\varphi$ constraints on variations of
intensities. Second, we have $\nu$ adsorption equations for the coexisting
interface phases. They contain an additional variable $\gamma$, which
we can eliminate and obtain $(\nu-1)$ equations of constraint. As
a result, the number of degrees of freedom is $(k+2)-\varphi-(\nu-1)$.
The interface phase rule becomes
\begin{equation}
f_{i}=k+3-\varphi-\nu.\label{eq:48}
\end{equation}

From this rule, we can find the maximum possible number of coexisting
interface phases (when $f_{i}=0$):
\begin{equation}
\nu_{max}=k+3-\varphi.\label{eq:49}
\end{equation}
For example, in a single-component two-phase system, only two interface
phases can co-exist at the interface. At a triple point ($\varphi=3$),
only one phase can exist at each interface. On the other hand, if
we have a binary single-phase system, an interface such as a surface,
can support up to four coexisting interface phases.

\section{Line phases}

\subsection{Definition of a line phase}

In the analysis of coexisting interface phases, we have so far neglected
the contributions of the line defects lying in the interface plane
and separating interface phases. Such line defects, or simply \emph{lines},
are 1D analogs of phase boundaries in bulk thermodynamics. We are
not aware of experimental studies of lines, but they have recently
been observed in atomistic computer simulations of metallic grain
boundaries.\cite{Frolov2013} Fig.~\ref{fig:Phase-transformation}(a)
shows an example of a straight line separating two grain boundary
phases, called kites and split kites, in a symmetrical tilt boundary
in Cu. The structure of this grain boundary undergoes a reversible
transformation from one phase to another with temperature. At some
temperature, the two phases coexist in equilibrium and are separated
by a line. It is also possible to equilibrate and isolated inclusion
of the split-kite phase bounded by a curved line (Fig.~\ref{fig:Phase-transformation}(b)).
The oval shape of the line indicates that its properties are anisotropic.
These observations motivate the development of a thermodynamic theory
of lines. Similar to bulk and interface phases, one can expect that
lines may exist in multiple phases. The goal of this section is to
examine thermodynamic properties of line phases. It should be noted
that line phases were not discussed by Gibbs\cite{Willard_Gibbs}
or analyzed by other researchers after Gibbs. However, they are important
as they influence the nucleation of new interface phases.

\begin{figure}
\noindent \begin{centering}
\includegraphics[clip,width=0.85\textwidth]{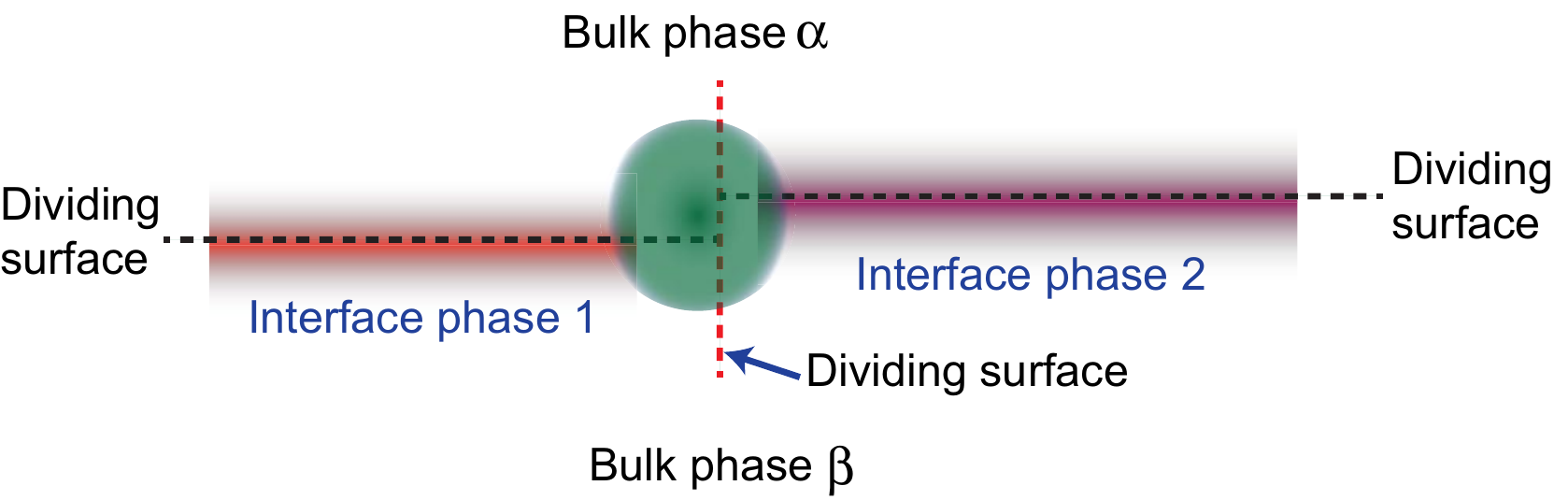}
\par\end{centering}

\protect\caption{Two interface phases separated by a line normal to the page. \label{fig:lines}}
\end{figure}

As a first step, we define excess properties of a line. Consider two
bulk phases $\alpha$ and $\beta$ and two interface phases 1 and
2 joining along a line as in Fig.~\ref{fig:lines}. We first choose
a geometric dividing surface for each interface phase as we did before;
for example, it can be the equimolar surface of component 1. This
choice is unimportant because it will later be generalized to arbitrary
excesses. Next, we choose a geometric dividing plane between the interface
phases. This plane is normal to the interface and parallel to the
line, but otherwise is arbitrary. For any extensive property $X$,
we define its line excess $\hat{X}$ as the total amount of $X$ in
the system, minus the bulk values of $X$ as if the bulk phases were
homogeneous all the way to the dividing surfaces, and minus the excesses
for the interface phases computed as if these phases remained homogeneous
(over the interface plane) all the way to the dividing plane: 
\begin{equation}
\hat{X}\equiv X-X_{\alpha}-X_{\beta}-\tilde{X}_{1}-\tilde{X}_{2}.\label{eq:50}
\end{equation}
For definitiveness, let us choose the dividing plane so that the line
excess of component 1 is zero, $\hat{N_{1}}=0$.

We now define a line phase by direct analogy with the previous definitions
of bulk and interface phases.

\textsl{\uline{Definition.}} Line phase is a set of spatially homogeneous
(along the line) states of a line described by a given fundamental
equation 
\begin{equation}
\hat{S}=\hat{S}(\hat{U},L,\hat{N}_{2},...,\hat{N}_{k})\label{eq:51}
\end{equation}
 with the following properties:
\begin{itemize}
\item $(\hat{S},\hat{U},\hat{N}_{2},...,\hat{N}_{k})$ are extensive (additive)
parameters along the line
\item $\hat{S}(\hat{U},L,\hat{N}_{2},...,\hat{N}_{k})$ is a homogeneous
function of first degree with respect to the variable set $(\hat{U},L,\hat{N}_{1},...,\hat{N}_{k})$ 
\item $\hat{S}(\hat{U},L,\hat{N}_{2},...,\hat{N}_{k})$ is a smooth (infinitely
differentiable) function
\end{itemize}
Here, $L$ is the length of the line. The only difference between
this definition and the previous definition of an interface phase
is that the spatial homogeneity is understood in the 1D sense (along
the line). As with bulk and interface phases, it is assumed that the
line phase follows its fundamental equation (\ref{eq:51}) even when
it is not in equilibrium with the surrounding phases.

\subsection{Coexistence of line phases}

To describe heterogeneous lines, we introduce the following postulates:

\textsl{\uline{Postulate 1.}} Any homogeneous line can potentially
exist in multiple line phases, each with its own fundamental equation. 

\textsl{\uline{Postulate 2.}} Any inhomogeneous line is composed
of homogeneous segments representing line phases.

As before, the homogeneous segments mentioned in Postulate 2 can be
either different line phases or different states of the same line
phase. 

We can now formulate the conditions of thermodynamic equilibrium in
a system composed of two bulk phases $\alpha$ and $\beta$, two interface
phases 1 and 2, and several line phases. Let the total number of line
phases be $\omega$. The total entropy of the system is obtained by
summing up the relevant fundamental equations:
\begin{eqnarray}
S_{tot} & = & S_{\alpha}(U_{\alpha},V_{\alpha},N_{\alpha1},...,N_{\alpha k})+S_{\beta}(U_{\beta},V_{\beta},N_{\beta1},...,N_{\beta k})\nonumber \\
 & + & \tilde{S}_{1}(\tilde{U}_{1},A_{1},\tilde{N}_{12},...,\tilde{N}_{1k})+\tilde{S}_{2}(\tilde{U}_{2},A_{2},\tilde{N}_{22},...,\tilde{N}_{2k})\nonumber \\
 & + & \sum_{n=1}^{\omega}\hat{S}_{n}(\hat{U}_{n},L_{n},\hat{N}_{n2},...,\hat{N}_{nk}).\label{eq:52}
\end{eqnarray}
In an isolated system, variations of $S_{tot}$ are subject to the
constraints of fixed total energy, total volume, total area of the
interface phases, total length of the line phases, and the total amount
of each chemical component:
\begin{equation}
U_{\alpha}+U_{\beta}+\tilde{U}_{1}+\tilde{U}_{2}+\sum_{n=1}^{\omega}\hat{U}_{n}=\textrm{const},\label{eq:53}
\end{equation}
\begin{equation}
V_{\alpha}+V_{\beta}=\textrm{const},\label{eq:54}
\end{equation}
\begin{equation}
A_{1}+A_{2}=\textrm{const},\label{eq:55}
\end{equation}
\begin{equation}
\sum_{n=1}^{\omega}L_{n}=\textrm{const},\label{eq:56}
\end{equation}
\begin{equation}
N_{\alpha i}+N_{\beta i}+\tilde{N}_{1i}+\tilde{N}_{2i}+\sum_{n=1}^{\omega}\hat{N}_{ni}=\textrm{const},\enskip i=1,...,k.\label{eq:57}
\end{equation}
Imposing these constraints by appropriate Lagrange multipliers, the
necessary conditions of $S_{tot}\rightarrow max$ are summarized as
follows:
\begin{itemize}
\item Thermal equilibrium
\begin{equation}
T_{\alpha}=T_{\beta}=\left(\dfrac{\partial\tilde{S}_{1}}{\partial\tilde{U}_{1}}\right)^{-1}=\left(\dfrac{\partial\tilde{S}_{2}}{\partial\tilde{U}_{2}}\right)^{-1}=\left(\dfrac{\partial\hat{S}_{1}}{\partial\hat{U}_{1}}\right)^{-1}=...=\left(\dfrac{\partial\hat{S}_{\omega}}{\partial\hat{U}_{\omega}}\right)^{-1}\equiv T.\label{eq:58}
\end{equation}

\item Mechanical equilibrium
\begin{equation}
p_{\alpha}=p_{\beta}.\label{eq:59}
\end{equation}

\item Chemical equilibrium
\begin{eqnarray}
\mu_{\alpha i}=\mu_{\beta i} & = & -T\left(\dfrac{\partial\tilde{S}_{1}}{\partial\tilde{N}_{1i}}\right)=-T\left(\dfrac{\partial\tilde{S}_{2}}{\partial\tilde{N}_{2i}}\right)\nonumber \\
 & = & -T\left(\dfrac{\partial\hat{S}_{1}}{\partial\hat{N}_{1i}}\right)=..=-T\left(\dfrac{\partial\hat{S}_{\omega}}{\partial\hat{N}_{\omega i}}\right)\equiv\mu_{i}\enskip i=2,...,k,\label{eq:60}
\end{eqnarray}
\begin{equation}
\mu_{\alpha1}=\mu_{\beta1}.\label{eq:61}
\end{equation}

\item Interface tension equilibrium
\begin{equation}
\dfrac{\partial\tilde{S}_{1}}{\partial A_{1}}=\dfrac{\partial\tilde{S}_{2}}{\partial A_{2}}.\label{eq:62}
\end{equation}

\item Line tension equilibrium
\begin{equation}
\dfrac{\partial\hat{S}_{1}}{\partial L_{1}}=...=\dfrac{\partial\hat{S}_{\omega}}{\partial L_{\omega}}.\label{eq:63}
\end{equation}

\end{itemize}
Equation (\ref{eq:58}) expresses the uniformity of temperature across
the equilibrium system, including the bulk, interface and line phases.
It is convenient to call the derivative $-T(\partial\hat{S}_{n}/\partial\hat{N}_{ni})$
the ``chemical potential'' of component $i$ in the line phase $n$.
Then the chemical equilibrium condition can be formulated as equality
of chemical potentials $\mu_{i}$ in the bulk, interface and line
phases. The interface tension equilibrium condition (\ref{eq:62})
reduces to $\gamma_{1}=\gamma_{2}$. Finally, defining the line tension
of a line phase $n$ by 
\begin{equation}
\tau_{n}=-T\dfrac{\partial\hat{S}_{n}}{\partial L_{n}},\label{eq:64}
\end{equation}
Eq.(\ref{eq:63}) states that the line tensions of coexisting line
phases must be equal: 
\begin{equation}
\tau_{1}=...=\tau_{\omega}.\label{eq:65}
\end{equation}
The line tension is obviously a 1D analog of the bulk pressure (up
to the sign) and the interface tension $\gamma$.

\subsection{The line adsorption equation}

Returning to a single line phase, consider reversible thermodynamic
processes in which the line remains in equilibrium with the interface
and bulk phases. Applying, as usual, the Euler theorem to the fundamental
equation (\ref{eq:51}) we have 
\begin{eqnarray}
\hat{S} & = & \dfrac{\partial\hat{S}}{\partial\hat{U}}\hat{U}+\dfrac{\partial\hat{S}}{\partial L}L+\sum_{i=2}^{k}\dfrac{\partial\hat{S}}{\partial\hat{N}_{i}}\hat{N}_{i}\nonumber \\
 & = & \dfrac{1}{T}\hat{U}-\dfrac{\tau}{T}L-\dfrac{1}{T}\sum_{i=2}^{k}\mu_{i}\hat{N}_{i},\label{eq:66}
\end{eqnarray}
from which
\begin{equation}
\tau L=\hat{U}-T\hat{S}-\sum_{i=1}^{k}\mu_{i}\hat{N}_{i}.\label{eq:67}
\end{equation}

On the other hand, differentiation of the fundamental equation (\ref{eq:51})
gives
\begin{equation}
d\hat{S}=\dfrac{1}{T}d\hat{U}-\dfrac{\tau}{T}dL-\dfrac{1}{T}\sum_{i=2}^{k}\mu_{i}d\hat{N}_{i},\label{eq:68}
\end{equation}
and thus
\begin{equation}
d\hat{U}=Td\hat{S}+\sum_{i=2}^{k}\mu_{i}d\hat{N}_{i}+\tau dL.\label{eq:69}
\end{equation}
Finally, adding Eq.(\ref{eq:69}) to the differential of Eq.(\ref{eq:67})
we obtain the Gibbs adsorption equation for a line phase:
\begin{equation}
Ld\tau=-\hat{S}dT-\sum_{i=2}^{k}\hat{N}_{i}d\mu_{i}.\label{eq:70}
\end{equation}
Equations (\ref{eq:66}) to (\ref{eq:70}) bear a close similarity
with the respective equations (\ref{eq:35}) to (\ref{eq:39}) of
interface thermodynamics. This is not surprising given the similar
structures of the fundamental equations (\ref{eq:21}) and (\ref{eq:51})
defining the interface and line phases, respectively.

\subsection{Reformulation in generalized line excesses}

The $k$ differentials appearing in the line adsorption equation (\ref{eq:70})
are not all independent because we have not yet imposed the condition
of equality of the surface tensions of the interface phases separated
by the line. The latter condition reduces the number of independent
differentials to $(k-1)$. To express the adsorption equation in terms
of independent differentials, we will reformulate it in terms of generalized
excesses.

\begin{figure}
\noindent \begin{centering}
\includegraphics[clip,width=0.67\textwidth]{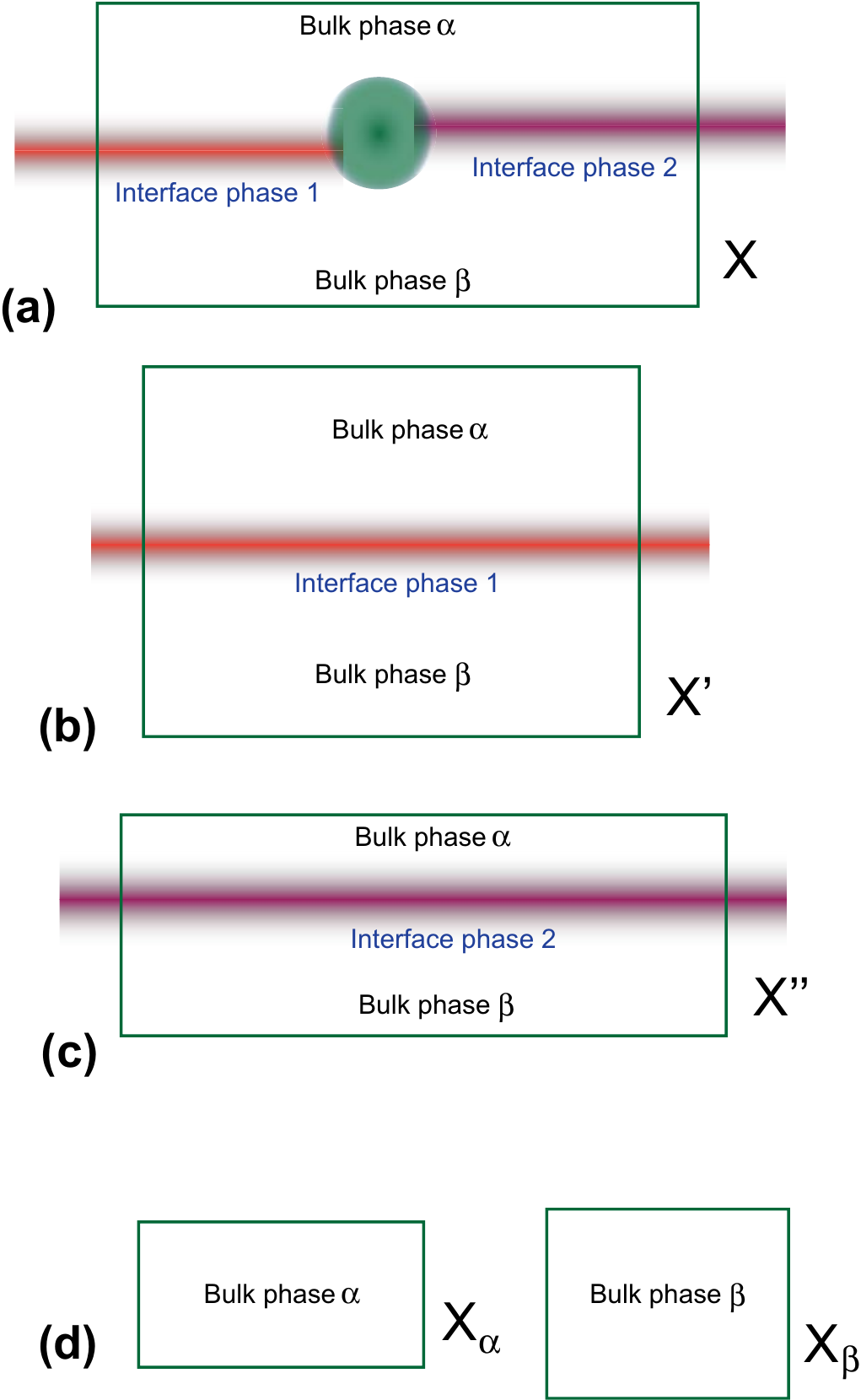}
\par\end{centering}

\protect\caption{Reference boxes used for the calculation of line excess properties.
(a) Two interface phases separated by a line. (b) Interface phase
1. (c) Interface phase 2. (d) Bulk phases $\alpha$ and $\beta$.
The values of an extensive property $X$ are indicated next to the
boxes. \label{fig:Reference-boxes}}
\end{figure}

The first step is to ``unwrap'' Eq.(\ref{eq:70}) by replacing all
excess quantities by their definitions (\ref{eq:50}). After rearrangements
with the aid of Eqs.(\ref{eq:7}) and (\ref{eq:39}), we obtain the
global form of the line adsorption equation:
\begin{equation}
Ld\tau=-SdT+Vdp-\sum_{i=1}^{k}N_{i}d\mu_{i}-Ad\gamma.\label{eq:71}
\end{equation}
Here, the quantities $S$, $V$, $N_{i}$ and $A$ refer to an arbitrarily
chosen rectangular box containing the two interface phases and the
line (Fig.~\ref{fig:Reference-boxes}(a)). The dimensions of the
box must be much larger than the characteristic thickness of the interface
and the cross-section of the line. Equation (\ref{eq:71}) shows that
$\tau$ is independent of definitions of excesses. It must be supplemented
by four other equations containing the same differentials, namely,
the global forms of the adsorption equations for the interface phases,
\begin{equation}
0=-S^{\prime}dT+V^{\prime}dp-\sum_{i=1}^{k}N_{i}^{\prime}d\mu_{i}-A^{\prime}d\gamma,\label{eq:72}
\end{equation}
\begin{equation}
0=-S^{\prime\prime}dT+V^{\prime\prime}dp-\sum_{i=1}^{k}N_{i}^{\prime\prime}d\mu_{i}-A^{\prime\prime}d\gamma,\label{eq:73}
\end{equation}
and the Gibbs-Duhem equations for the bulk phases,
\begin{equation}
0=-S_{\alpha}dT+V_{\alpha}dp-\sum_{i=1}^{k}N_{\alpha i}d\mu_{i},\label{eq:74}
\end{equation}
\begin{equation}
0=-S_{\beta}dT+V_{\beta}dp-\sum_{i=1}^{k}N_{\beta i}d\mu_{i}.\label{eq:75}
\end{equation}
Equations (\ref{eq:72}) and (\ref{eq:73}) are written for imaginary
boxes containing a single interface phase, either 1 or 2, and uninfluenced
by the line (Fig.~\ref{fig:Reference-boxes}(b,c)). The cross-sectional
areas and the total amounts of extensive properties in these boxes
are distinguished by the prime and double prime, respectively. Likewise,
Eqs.(\ref{eq:74}) and (\ref{eq:75}) represent homogeneous bulk regions
of phases $\alpha$ and $\beta$, respectively (Fig.~\ref{fig:Reference-boxes}(d)).
Because the differential coefficients in each of the Eqs.(\ref{eq:71})
to (\ref{eq:75}) can be scaled by an arbitrary common multiplier,
the choice of dimensions of all boxes is arbitrary as long as the
conditions stated above (e.g., the homogeneity of the bulk regions)
are satisfied.

The right-hand side of Eq.(\ref{eq:71}) contains $(k+3)$ terms.
Equations (\ref{eq:72}) to (\ref{eq:75}) impose four constraints,
leaving $(k-1)$ independent differentials as the variables of $\tau$.
Which variables to eliminate is a matter choice and can be conveniently
controlled by re-writing Eq.(\ref{eq:71}) in terms of generalized
excess as it was done for interfaces (sec.~\ref{sub:Reformulation-interfaces}).
Applying the Kramer rule of linear algebra,\cite{Cahn79a} we obtain
\begin{equation}
Ld\tau=-\left[S\right]_{WXYZ}dT+\left[V\right]_{WXYZ}dp-\sum_{i=1}^{k}\left[N_{i}\right]_{WXYZ}d\mu_{i}-\left[A\right]_{WXYZ}d\gamma,\label{eq:76}
\end{equation}
where $W$, $X$, $Y$ and $Z$ are any four of the extensive variables
$(S,V,N_{1},...,N_{k},A)$. The square brackets are generalized excesses
defined by ratios of two determinants of ranks 5 and 4. Namely, for
any extensive property $R$, 
\begin{equation}
\left[R\right]_{WXYZ}\equiv\dfrac{\left\vert \begin{array}{ccccc}
R & W & X & Y & Z\\
R^{\prime\prime} & W^{\prime\prime} & X^{\prime\prime} & Y^{\prime\prime} & Z^{\prime\prime}\\
R^{\prime} & W^{\prime} & X^{\prime} & Y^{\prime} & Z^{\prime}\\
R_{\alpha} & W_{\alpha} & X_{\alpha} & Y_{\alpha} & Z_{\alpha}\\
R_{\beta} & W_{\beta} & X_{\beta} & Y_{\beta} & Z_{\beta}
\end{array}\right\vert }{\left\vert \begin{array}{cccc}
W^{\prime\prime} & X^{\prime\prime} & Y^{\prime\prime} & Z^{\prime\prime}\\
W^{\prime} & X^{\prime} & Y^{\prime} & Z^{\prime}\\
W_{\alpha} & X_{\alpha} & Y_{\alpha} & Z_{\alpha}\\
W_{\beta} & X_{\beta} & Y_{\beta} & Z_{\beta}
\end{array}\right\vert }.\label{eq:77}
\end{equation}
The meaning of $\left[R\right]_{WXYZ}$ is the line excess of property
$R$ computed with a set of reference boxes such that the excesses
of $W$, $X$, $Y$ and $Z$ are zero. Equation (\ref{eq:76}) reveals
two new excess quantities that did not appear in Eq.(\ref{eq:70}):
the line excess volume $\left[V\right]_{WXYZ}$ characterizing the
contribution of the line to the total interface excess volume, and
the line excess area $\left[A\right]_{WXYZ}$ characterizing the interface
area attributed to the line. Equation (76) shows that a line tension
depends on both the 3D pressure $p$ and the 2D ``pressure'' $\gamma$.
The line excess volume and area, as well as the line excess entropy
$\left[S\right]_{WXYZ}$ and the line segregations $\left[N_{i}\right]_{WXYZ}$,
are not unique. Their values depend on the choice of the properties
$W$, $X$, $Y$ and $Z$. Whatever their choice is, the terms with
the excesses of $W$, $X$, $Y$ and $Z$ disappear and we are left
with $(k-1)$ independent differentials. For example, by choosing
$WXYZ=ASVN_{1}$, the line adsorption equation reduces to 
\begin{equation}
Ld\tau=-\sum_{i=2}^{k}\left[N_{i}\right]_{ASVN_{1}}d\mu_{i},\label{eq:78}
\end{equation}
involving only line segregations. 

The excess form of the line tension, which was previously given by
Eq.(\ref{eq:67}), can also be generalized to 
\begin{equation}
\tau L=\left[U\right]_{WXYZ}-T\left[S\right]_{WXYZ}+p\left[V\right]_{WXYZ}-\sum_{i=1}^{k}\mu_{i}\left[N_{i}\right]_{WXYZ}-\left[A\right]_{WXYZ}\gamma.\label{eq:79}
\end{equation}
For example, the choice of $WXYZ=ASVN_{1}$ reduces this equation
to
\begin{equation}
\tau L=\left[U-\sum_{i=2}^{k}\mu_{i}N_{i}\right]_{ASVN_{1}}.\label{eq:80}
\end{equation}

As an application of Eq.(\ref{eq:76}), consider two coexisting line
phases that remain in equilibrium with each other during a reversible
process. Writing Eq.(\ref{eq:76}) for each phase, subtracting these
equations and taking into account that the differentials $d\tau$
remain equal for both phases, we obtain
\begin{eqnarray}
 & - & \Delta\left(\dfrac{\left[S\right]_{WXYZ}}{L}\right)dT+\Delta\left(\dfrac{\left[V\right]_{WXYZ}}{L}\right)dp-\sum_{i=1}^{k}\Delta\left(\dfrac{\left[N_{i}\right]_{WXYZ}}{L}\right)d\mu_{i}\nonumber \\
 & - & \Delta\left(\dfrac{\left[A\right]_{WXYZ}}{L}\right)d\gamma=0.\label{eq:81}
\end{eqnarray}
Here, the symbol $\Delta$ designates the difference between the excess
properties (per unit length) of the two line phase. This equation
describes a $(k-2)$-dimensional phase coexistence hypersurface in
the $(k-1)$-dimensional space of variables and is analogous to the
Clapeyron-Clausius equation for coexisting bulk systems. For example,
for $WXYZ=ASVN_{1}$ the coexistence hypersurface exists is the space
of chemical potentials $(\mu_{2},...,\mu_{k})$ and is described by
the equation
\begin{equation}
\sum_{i=2}^{k}\Delta\left(\dfrac{\left[N_{i}\right]_{WXYZ}}{L}\right)d\mu_{i}=0,\label{eq:82}
\end{equation}
where the differential coefficients are the jumps in line segregations
(per unit length) across the hypersurface.

\subsection{The line phase rule}

Consider an equilibrium system containing $\varphi$ bulk phases that
we want to keep during all reversible variations of parameters. We
single out one particular interface containing $\nu$ interface phases
and consider a particular line between two interface phases 1 and
2. Suppose this line contains $\omega$ line phases. How many variables
can we vary while keeping all these phases in coexistence? 

Out of the $(k+2)$ intensities of our system, $\varphi$ can be eliminated
by the Gibbs-Duhem equations for the bulk phases. The adsorption equations
for the interface phases impose $\nu$ more constraints. The adsorption
equations for the line phases add another $\omega$ constraints. However,
the interface and line adsorption equations contain the additional
differentials $d\gamma$ and $d\tau$. Eliminating them, the total
number of constraints becomes $(\varphi+\nu+\omega-2)$. Thus, the
number of remaining degrees of freedom is
\begin{equation}
f_{L}=k+4-\varphi-\nu-\omega.\label{eq:83}
\end{equation}
This equation is the phase rule for line phases. The maximum possible
number of coexisting line phases (when $f_{L}=0$) is
\begin{equation}
\omega_{max}=k+4-\varphi-\nu.\label{eq:84}
\end{equation}

If we only concerned with two bulk phases separated by an interface
containing one line, then $\varphi=2$, $\nu=2$, and the phase rule
reduces to $f_{L}=k-\omega$. Accordingly, the maximum number of coexisting
line phases is $\omega_{max}=k$.

\section{Discussion}

Table \ref{tab:phase-rules} summarizes the phase rules derived here
for the bulk, interface and line phases. The last column contains
the maximum number of phases that can coexist in equilibrium. All
these phase rules can be summarized in the equations 

\begin{equation}
f=k+5-d-\theta,\label{eq:85}
\end{equation}
\begin{equation}
\theta_{max}=k+5-d,\label{eq:86}
\end{equation}
where $d$ is the smallest dimensionality of phases included into
consideration and $\theta=\varphi+\nu+\omega$ is the number of coexisting
phases in the system. For a given $d$, the number of phases of a
lower dimensionality must be excluded from $\theta$. For example,
$\omega=0$ for interface phases (Table \ref{tab:phase-rules}).

These equations can be used to predict the number of independent variables
and the maximum possible number of coexisting phases, depending on
the dimensionality of the phases. For example, in a binary system,
a phase boundary can support a maximum of $\nu_{max}=3$ coexisting
interface phases ($d=2$, $k=2$, $\varphi=2$, $\omega=0$). If pressure
is fixed, then only two. In a single-component system, an interface
can support only two interface phases and the line separating them
can have only one line phase ($d=1$, $k=1$, $\varphi=2$, $\nu=2$). 

In the foregoing discussion, we assumed that interfaces and lines
separated distinct phases. This analysis is readily extended to single-phase
interfaces such as grain boundaries and lines separating regions of
the same interface phase with different crystallographic orientations.
In such cases, the choice of the dividing surface is arbitrary and
all excesses $\tilde{N}_{1},...,\tilde{N}_{k}$ (accordingly, $\hat{N}_{1},...,\hat{N}_{k}$)
must appear as arguments in the fundamental equations. All calculations
remain similar and lead to the same phase rules with an appropriate
count of phases (e.g., $\varphi=1$ for a grain boundary).

Using the same thermodynamic approach, it is straightforward to derive
a phase rule for 0-dimensional defects formed between neighboring
line phases (Fig.~\ref{fig:interface_phases}(b)), which could be
called ``points''. 

The analysis presented here was based on a simplified treatment and
its immediate applications are restricted to multicomponent fluids.
This treatment is not applicable to solid phases without appropriate
modifications.\cite{Larche73,Larche_Cahn_78,Larche1985,Frolov2010d,Frolov2010e,Frolov2012a,Frolov2012b}
Interfaces in solid systems are often anisotropic and their properties
may depend on the crystallographic orientation of the interface plane.
In terms of Cahn's classification,\cite{Cahn82a} all interface phase
transformations considered here are congruent. Furthermore, we identified
the interface free energy with the interface stress and referred to
this common property as ``tension''. While this is correct for fluid
systems, for solid-solid interfaces the interface free energy and
interface stress are different quantities both conceptually and numerically.\cite{Willard_Gibbs,Shuttleworth50,Cahn79a,Kramer07,Frolov2010d,Frolov2010e,Frolov2012a,Frolov2012b}
A line phase can also be characterized by a line free energy and a
line stress, which are different properties. We neglected chemical
reactions in the bulk or at interfaces and lines. Finally, for the
sake of simplicity we neglected the curvature effects on the interface
and line properties. The goal of the present work was to demonstrate
the general approach and outline the direction of future work. Extensions
of the present analysis to include the effects mentioned above are
possible and would result in thermodynamic theories and phases rules
for low-dimensional phases in a wider range of real materials. The
nucleation of new interface phases is not well understood or theoretically
described. Developing a more detailed thermodynamic theory of lines
is the first necessary step in this direction. Much can be done in
this field.

It should be clarified that the proposed approach does not ignore
the fundamental differences in physical properties of low-dimensional
versus bulk phases. In fact, the phase rules (\ref{eq:85}) and (\ref{eq:86})
derived here contain the dimensionality of the system as a parameter.
Bulk phases, interfaces and lines belong to different universality
classes and exhibit different critical behaviors (e.g., critical exponents),
as well as many other physical properties.\cite{Forgacs:1991aa} However,
the basic thermodynamic formalism and the rules for the identification
of independent thermodynamic variables and description of phase equilibria
remain exactly the same for any dimensionality of space. In its postulational
basis, thermodynamics is blind to the dimensionality of space. The
postulates of thermodynamics are formulated in abstract concepts such
as a system, a state, a variable, extensive and intensive parameters,
conservation, etc.\cite{Tisza:1961aa} that do not involve the real
space or its dimensionality. As a consequence, the fundamental equation
of any phase has the same mathematical structure regardless of whether
the phase exists in 3D space, at an interface or in a line defect.
This explains the remarkable similarity, in fact identity, in the
phase equilibrium descriptions for the bulk and low-dimensional systems.

While the concepts of 2D phases and 2D phase diagrams have long been
accepted and successfully used in the surface/interface physics and
chemistry communities and were later adopted in materials science,\cite{Hart:1972aa,Cahn82a,Rottman1988a}
a recent trend in the materials community is to reject the terms ``interface
phase'' and ``2D phase'' on the ground that such phases ``do not
satisfy the Gibbs definition of a phase''.\cite{Tang06,Tang06b,Dillon2007,Kaplan2013,Cantwell-2013}
It is pointed out that they do not meet Gibbs' requirement of homogeneity
and in addition cannot exist without being in contact with bulk phases.
Gibbs' definition of a phase\cite{Willard_Gibbs} was discussed in
Sec.~\ref{sub:bulk_phase}. Mathematically, his requirement of homogeneity
is expressed by the homogeneity of first degree of the fundamental
equation of the phase. For interfaces, the fundamental equation is
homogeneous with respect to the area, and for lines with respect to
the length. In other words, the homogeneity of a phase is embedded
in its definition (\ref{eq:0}) for any dimensionality. The requirement
that we should be able to physically extract any given phase from
the rest of the system is not part of Gibbs' thermodynamics.\cite{Willard_Gibbs}
Is it not part of the modern logical structure of thermodynamics either,
nor is it needed for any thermodynamic derivations involving phases.
As long as a particular part of a system follows its own fundamental
equation satisfying the mathematical properties stated above and can
exchange extensive properties with the rest of the system, it satisfies
the thermodynamic definition of a phase.

\section{Conclusions}

To summarize, we have presented a unified thermodynamic description
of phases and phase equilibria in 3D, 2D and 1D systems. In all cases,
the phase is identified with a fundamental equation of state, see
Eq.(\ref{eq:0}). The fundamental equation defines a phase and encapsulates
all of its properties. In all dimensions, the phases are treated the
same way and are described by similar thermodynamic equations. The
same thermodynamic formalism can be applied for the description of
phase equilibria and phase transformations in bulk systems, interfaces
and line defects separating 2D interface phases. For both lines and
interfaces, we have rigorously derived adsorption equations in terms
of generalized excess quantities. We have also derived phase coexistence
equations that can be utilized for the construction of phase diagrams
for low-dimensional systems. The Gibbs phase rule describing the coexistence
of bulk phases has been generalized to phase rules for interfaces
and lines. Such rules predict the number of thermodynamic degrees
of freedom and the maximum number of phases than can coexist in the
systems of the respective dimensionality.

\begin{figure}
\noindent \textbf{(a)}\includegraphics[clip,width=0.5\textwidth]{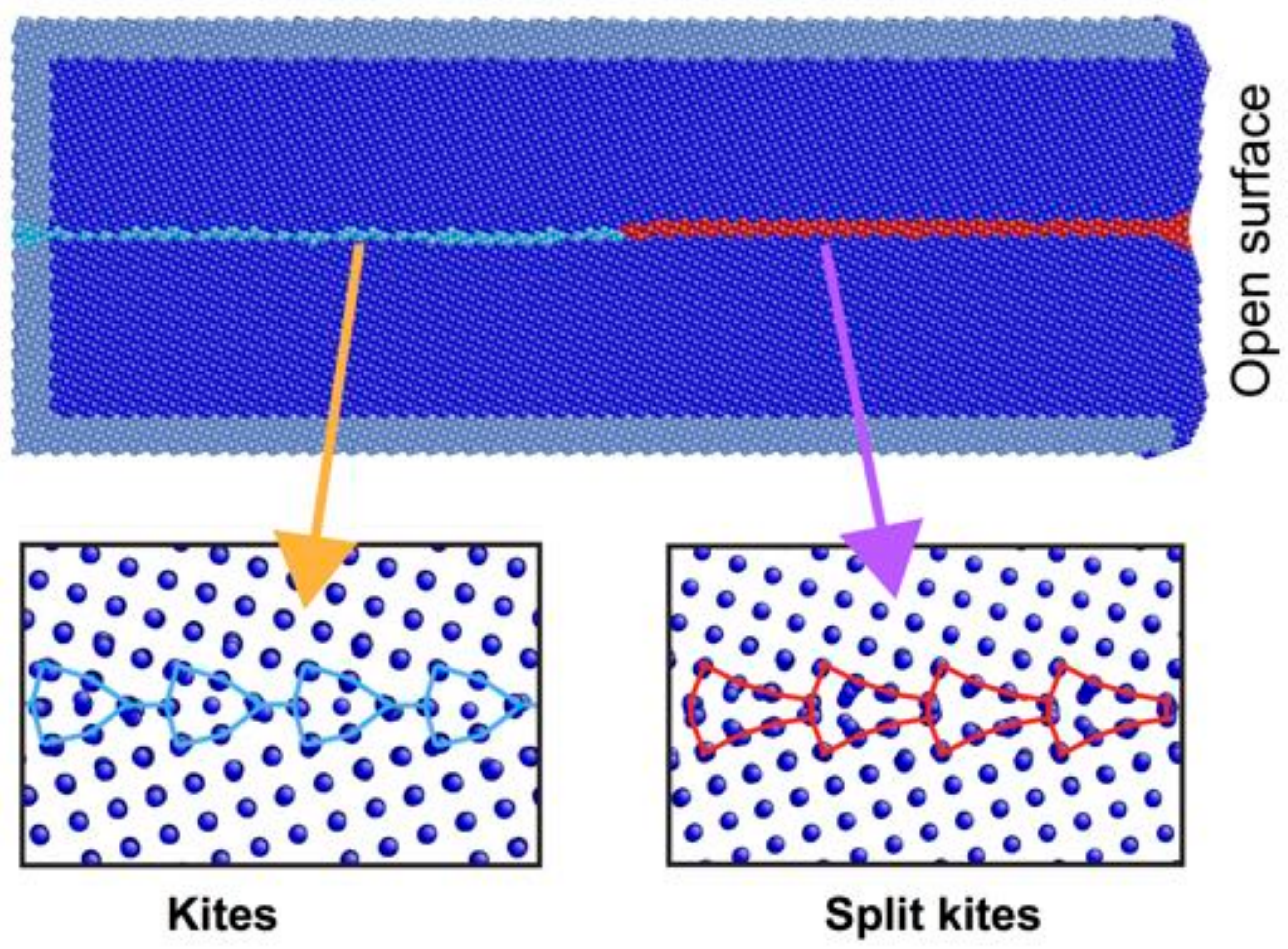}\bigskip{}

\noindent \textbf{(b)}\includegraphics[clip,width=0.5\textwidth]{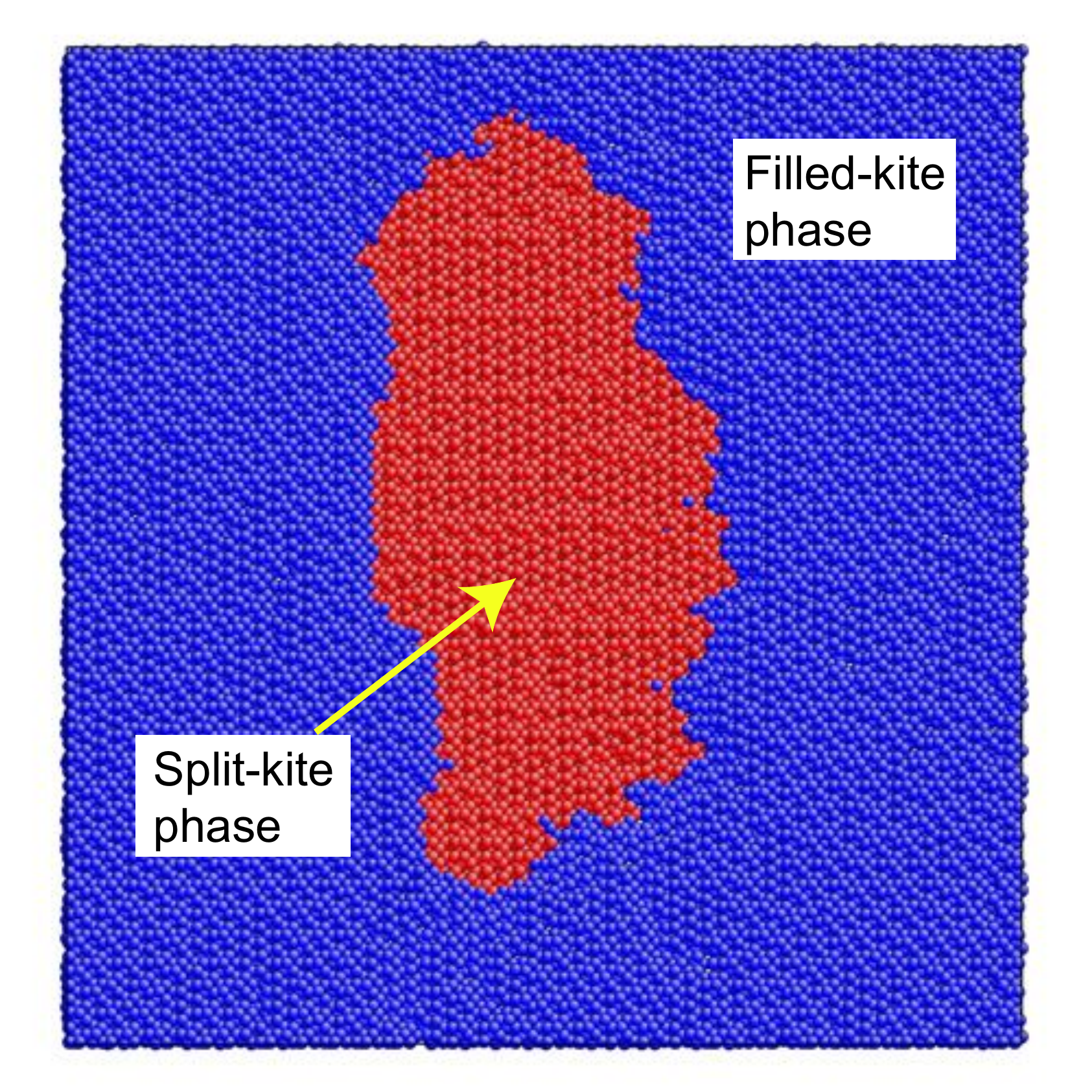}

\protect\caption{Examples of lines: 1D defects separating interface phases. (a) Phase
transformation in the Cu $\Sigma5\,(310)$ grain boundary at the temperature
of 1000 K.\cite{Frolov2013} The boundary was initially composed of
kite-shape structural units (light blue). A new phase composed of
split-kite structural units (red) grows from the surface and eventually
penetrates all through the sample. (b) Top view of a two-phase state
of the Cu $\Sigma5\,(210)$ grain boundary at the temperature of 700
K.\cite{Frolov_to_be_published_2015} An inclusion of a split-kite
phase is surrounded by a matrix of the filled-kite phase. The interface
phases are separated by an oval-shape line defect. \label{fig:Phase-transformation}}
\end{figure}

Recent years have seen a significant increase in the research activity
dedicated to interface phase transformations. Experiments have uncovered
a number GB phases with discrete thicknesses and various segregation
patterns in binary and multi-component metallic alloys\cite{Ma2012,Luo23092011,Cantwell-2013}
and ceramic materials.\cite{Dillon2007,Cantwell-2013,Kaplan2013}
It has been demonstrated that transformations among such phases can
strongly impact many engineering properties of materials such as grain
growth, mechanical behavior and interface transport.\cite{Divinski2012}
On the modeling side, atomistic simulations have revealed reversible
temperature-induced transformations between different structural phases
in metallic systems\cite{Frolov2013} and their effect on GB diffusion,\cite{Frolov2013a}
response to applied mechanical stresses\cite{Frolov:2014aa} and other
properties. Phases inside line defects have never been reported by
either experimentalists or modelers. However, we envision that such
phases may be discovered in the future. The present work predicts
their possible existence and describes the conditions of their thermodynamic
coexistence. It should be emphasized that lines play a critical role
in 2D phase transitions. For example, their excess free energy and
other properties determine the nucleation barriers of 2D phases as
illustrated by Fig.~\ref{fig:Phase-transformation}. 

At this juncture, it is important to develop theories capable of explaining
the experimental observations and simulation results and guiding new
research in this field. It is hoped that the present work contributes
to this course by providing a rigorous thermodynamic framework for
the description and prediction of phase equilibria in interfaces and
lines. As one example, the phase rules derived in this work provide
a guidance for phase diagram construction and design of new experiments
and simulations, as usually does the existing phase rule for bulk
systems. 

\vspace{0.15in}

\emph{Acknowledgements.} -- T.F. was supported by a post-doctoral
fellowship from the Miller Institute for Basic Research in Science
at the University of California, Berkeley. Y.M. was supported by the
National Science Foundation, Division of Materials Research, the Metals
and Metallic Nanostructures Program.


\newpage{}\clearpage{}

\begin{table}
\begin{tabular}{lccccc}
\hline 
 & Bulk phases & Interface phases & Line phases & $f$ & Maximum \# of phases\tabularnewline
\hline 
Bulk & $\varphi$ & 0 & 0 & $k+2-\varphi$ & $k+2$\tabularnewline
Interface & $\varphi$ & $\nu$ & 0 & $k+3-\varphi-\nu$ & $k+3-\varphi$\tabularnewline
Line & $\varphi$ & $\nu$ & $\omega$ & $k+4-\varphi-\nu-\omega$ & $k+4-\varphi-\nu$\tabularnewline
\hline 
\end{tabular}

\protect\caption{Summary of phase rules for bulk, interface and line phases. $k$ is
the number of chemical components, $\varphi$ is the number of bulk
phases, $\nu$ is the number of interface phases, $\omega$ is the
number of line phases, and $f$ is the number of degrees of freedom.\label{tab:phase-rules}}
\end{table}

\newpage{}\clearpage{}

\end{document}